\documentclass[12pt]{article}
\usepackage{amssymb} 
\usepackage{epsfig}
\newcommand{\be}{\begin{equation}}
\newcommand{\ee}{\end{equation}}
\newcommand{\bea}{\begin{eqnarray}}
\newcommand{\eea}{\end{eqnarray}}

\begin{document} 

\begin{center}
{\bf A LECTURE ON NEUTRINO MASSES, MIXING AND OSCILLATIONS}
\end{center}

\begin{center}
S. M. Bilenky 
\footnote {Report at the International School of Physics
``Enrico Fermi'', Varenna, August 2002.}\\

\end{center}
\vspace{0.1cm} 
\begin{center}

{\em INFN, Sez. di Torino and Dip. di Fisica Teorica,
Univ. di Torino, I-10125 Torino, Italy\\}

{\em  Joint Institute
for Nuclear Research, Dubna, R-141980, Russia\\}
\end{center}

\begin{abstract}
Neutrino mixing and basics of neutrino oscillations are 
considered.
Recent evidences in favour of neutrino oscillations, obtained in the solar and 
atmospheric neutrino experiments, are discussed. Neutrino oscillations 
in the solar and atmospheric ranges of 
$\Delta m^{2}$are considered in the framework of the minimal scheme with 
the mixing of three massive neutrinos.
\end{abstract}

\begin{center}

\section{Introduction}
\end{center}

There exist at present convincing evidences of neutrino oscillations
obtained in experiments with neutrinos from natural sources:
in the atmospheric \cite{S-K,Soudan, MACRO}
and in the solar neutrino experiments \cite{
Cl,GALLEX-GNO,SAGE,S-Ksol,SNO,SNONC,SNOCC}.
The observation of neutrino oscillations
give us  first evidence for nonzero neutrino masses and neutrino mixing.

The investigation of neutrino oscillations is based on:

\begin{enumerate}
\item 
Interaction of neutrinos with other particles is given by the Standard
Model. It was proved by numerous experiments, including very precise
LEP experiments, that 
the Standard Model perfectly describes 
experimental data in the energy region up to a few hundreds GeV.
The Standard Charged Current (CC) and Neutral Current (NC)
Lagrangians are given by
\be
\mathcal{L}_{I}^{\mathrm{CC}}
=
- \frac{g}{2\sqrt{2}} \,
j^{\mathrm{CC}}_{\alpha} \, W^{\alpha}
+
\mathrm{h.c.};
\,~~
\mathcal{L}_{I}^{\mathrm{NC}}
=
- \frac{g}{2\cos\theta_{W}} \,
j^{\mathrm{NC}}_{\alpha} \, Z^{\alpha}
\,.
\label{001}
\ee

Here $g$ is the SU(2)
gauge coupling constant,
$\theta_{W}$ is the weak angle, $W^{\alpha}$ and $Z^{\alpha}$
are fields of charged $W^{\pm}$ and neutral $Z^{0}$ vector bosons and 
for the leptonic charged current $j^{\mathrm{CC}}_{\alpha}$ and 
neutrino neutral current $j^{\mathrm{NC}}_{\alpha}$ we have

\begin{equation}
j^{\mathrm{CC}}_{\alpha} = \sum_{l} \bar \nu_{lL} \gamma_{\alpha}l_{L};\,~~
j^{\mathrm{NC}}_{\alpha} =\sum_{l} \bar \nu_{lL}\gamma_{\alpha}\nu_{lL}\,.
\label{002}
\end{equation}

\item Three flavour neutrinos $\nu_{e}$, $\nu_{\mu}$
and $\nu_{\tau}$ exist in nature.

From the LEP experiments on the measurement of the width of the  decay
 $Z \to \nu_{l} + \bar\nu_{l} $ for the number of flavour neutrinos $n_{\nu_{f}}$
it was obtained the value \cite{PDG}

\begin{equation}
n_{\nu_{f}} = 3.00 \pm 0.06 \,.
\label{003}
\end{equation}

From the global fit of the LEP data for $n_{\nu_{f}}$ it was found

\begin{equation}
n_{\nu_{f}} = 2.984 \pm 0.008 \,.
\label{004}
\end{equation}

\end{enumerate}

\begin{center}
\section{Neutrino mixing}
\end{center}

The hypothesis of neutrino mixing is based on the assumption
that there is {\em a neutrino mass term} in the total Lagrangian.
It was proposed several mechanisms of the generation of the neutrino mass term.
Later we will discuss the most popular see-saw mechanism \cite{see-saw}.

There are two types of possible neutrino mass terms (see \cite{BPet,BGG})

\begin{enumerate}
\item

Dirac mass term
\be
\mathcal{L}^{\mathrm{D}}=- \bar \nu'_{R}\,M^{\mathrm{D}} \nu'_{L}
+\mathrm{h.c.}
\label{005}
\ee
Here

\bea
\nu'_{L}=\left(
\begin{array}{c}
\nu_{e L}\\
\nu_{\mu L}\\
\nu_{\tau L}\\
\vdots
\end{array}
\right);\,~
\nu'_{R}=\left(
\begin{array}{c}
\nu_{e R}\\
\nu_{\mu R}\\
\nu_{\tau R}\\
\vdots
\end{array}
\right)
\label{006}
\eea

and $ M^{\mathrm{D}}$ is a complex non-diagonal matrix.

In the case of the Dirac mass term
the total Lagrangian is invariant under global gauge transformation

$$\nu'_{L}\to e^{i\,\alpha}\,\nu'_{L};\, \nu'_{R}\to e^{i\,\alpha}\,\nu'_{R};\,
l\to e^{i\,\alpha}\,l $$

This invariance means that the total lepton number 
$ L= \sum_{l}\,L_{l}$ is conserved.

\item

Majorana mass term

\be
\mathcal{L}^{\mathrm{Mj}}=-\frac{1}{2}\,( \overline{\nu'_{L}})^{c}\,M^{\mathrm{Mj}} \nu'_{L}
+\mathrm{h.c.}
\label{007}
\ee

Here $M^{\mathrm{Mj}}$ is a complex non-diagonal {\em symmetrical} matrix
and

$$(\nu'_{L})^{c}= C \bar\nu'^{T}_{L}\,,$$
where $C$ is the unitary matrix of the charge conjugation,
which satisfies the conditions
$C\,\gamma^{T}_{\alpha}\,C^{-1}= -\gamma_{\alpha};\,C^{T}= -C$.
It is obvious that in the case of the Majorana mass term
there are no any conserved lepton numbers.

\end{enumerate}

After the standard diagonalization of a neutrino mass term we have

\be
\nu_{lL} = \sum_{i} U_{li} \nu_{iL}\,,
\label{008}
\ee
where
$ U $ is a 
unitary mixing matrix and $\nu_{i}$
is the field of neutrino with mass $m_{i}$.

In the case of the Dirac mass  term $\nu_{i}$
is the field of the Dirac neutrinos and antineutrinos
which possess conserved lepton numbers ($L(\nu_{i})=1;\,L(\bar \nu_{i})=-1 $ ).
In the case of the Majorana mass term $\nu_{i}$
is the field of truly neutral Majorana neutrinos. The field $\nu_{i}$
satisfies the Majorana condition

\be
\nu_{i} =\nu^{c}_{i}= C \bar\nu^{T}_{i}.
\label{009}
\ee

If there are only 
flavour fields $\nu_{lL}$ 
in the column
$\nu'_{L}$, the number of the massive neutrinos
$\nu_{i}$ is equal to three
and $U$ is a 3$\times$ 3 
unitary matrix.

In the neutrino mass term it could be also fields, 
which do not enter into the standard CC and NC interactions.
Such fields are called sterile.
If in the coloumn $\nu'_{L}$ there are $n_{s}$ sterile 
fields $\nu_{s_{a}L}$, the number of massive neutrinos
$\nu_{i}$ is equal to $3+n_{s}$ and $U$ is a $(3+n_{s})\times (3+n_{s}) $ 
unitary matrix. In this case in addition to the mixing relation (\ref{008}) we have

\begin{equation}
\nu_{s_{a}L} = \sum_{i} U_{s_{a}i} \nu_{iL}\,.
\label{010}
\end{equation}
Sterile fields can be right-handed neutrino fields, SUSY fields etc. 
If more than three neutrino masses are small,    
transition of the flavour
neutrinos $\nu_{e}$, $\nu_{\mu}$, $\nu_{\tau}$ into sterile states
become possible.

If sterile fields are right-handed neutrino fields $\nu_{lR}$,
neutrino mass term has the form
of the sum of the left-handed Majorana, Dirac  and
right-handed Majorana mass terms: 

\bea
\mathcal{L}^{\mathrm{D+Mj}}&=&
-\frac{1}{2}\,( \overline{\nu_{L}})^{c}\,M^{\mathrm{Mj}}_{L}
 \nu_{L}-  \bar \nu_{R}\,M^{\mathrm{D}}\, \nu_{L}
-\frac{1}{2}\,\bar\nu_{R}\,M^{\mathrm{Mj}}_{R}
(\nu_{R})^{c} +\mathrm{h.c.} \nonumber\\ 
&=&-\frac{1}{2}\,( \overline{\nu'_{L}})^{c}\,M^{\mathrm{D+Mj}} \nu'_{L}
+\mathrm{h.c.}
\label{011}
\eea

Here 

\bea
\nu_{L}=\left(
\begin{array}{c}
\nu_{e L}\\
\nu_{\mu L}\\
\nu_{\tau L}
\end{array}
\right),\,~
\nu_{R}=\left(
\begin{array}{c}
\nu_{e R}\\
\nu_{\mu R}\\
\nu_{\tau R}
\end{array}
\right)\,,
\label{012}
\eea
$M^{\mathrm{Mj}}_{L}$ and $M^{\mathrm{Mj}}_{R}$
are complex non-diagonal symmetrical 3$\times$3 Majorana matrices and
$M^{\mathrm{D}}$ is a complex non-diagonal 3$\times$3 Dirac matrix.
The mass term $\mathcal{L}^{\mathrm{D+Mj}}$
is called the Dirac and Majorana mass term.
After the diagonalization of the mass term (\ref{011})
we have

\begin{equation}
\nu_{lL} = \sum_{i=1}^{6} U_{li} \nu_{iL};\,~
(\nu_{lR})^{c} = \sum_{i=1}^{6} U_{\bar l i} \nu_{iL},
\label{013}
\end{equation}
where $U$ is the unitary 6$\times$6 mixing matrix.

We will discuss now
the see-saw mechanism 
of neutrino mass generation \cite{see-saw}.
In order to explain an idea of the mechanism we will consider the simplest case of one
type of neutrino. Let us assume that the standard Higgs mechanism with one Higgs doublet, which is the mechanism of the generation of the masses of quarks and leptons, generates 
the Dirac neutrino mass term
\be
\mathcal{L}^{\mathrm{D}} = -m\,\bar \nu_{R}\nu_{L} +\mathrm{h.c.} 
\label{014}
\ee

It is natural to expect that the mass $m$ is of the same order of magnitude
as masses
of the corresponding lepton or quark. We know, however,
 from experimental data that
neutrino masses are much smaller than the masses of leptons and quarks.
In order to ``suppress'' neutrino mass let us assume that there exists
lepton number violating beyond the SM mechanism of the generation
of the right-handed Majorana mass term

\be
\mathcal{L}^{\mathrm{Mj}}_{R} = -M\,\bar \nu_{R}(\nu_{R})^{c} +\mathrm{h.c.}, 
\label{015}
\ee

with $M \gg m$ (usually it is assumed that $M \simeq M_{\rm{GUT}}\simeq
10^{15}\, \rm{GeV}$).

The total mass term is  the Dirac and Majorana one  
with

\bea
M^{\rm{D+Mj}}=\left(
\begin{array}{cc}
0&m\\
m&M
\end{array}
\right);\,~
\nu'_{L}=\left(
\begin{array}{c}
\nu_{L}\\
(\nu_{ R})^{c}
\end{array}
\right)
\label{016}
\eea

After the diagonalization of the mass term we have

\bea
\nu_{L} &=& i\,\cos \theta \,
\nu_{1L} +\sin \theta\, \nu_{2L}\nonumber\\ 
(\nu_{R})^{c} &=& -i\,\sin \theta \,\nu_{1L} +\cos \theta\, \nu_{2L}, 
\label{017}
\eea

where $\nu_{1}$ and $\nu_{2}$
are fields of the neutrino Majorana with masses

\be
m_{1}= -\frac{1}{2}\,+\frac{1}{2}\,\sqrt{M^{2}+ 4\,m^{2}}\simeq \frac{m^{2}}{M}\ll 1 
\label{018}
\ee

and

\be
m_{2}= \frac{1}{2}\,+\frac{1}{2}\,\sqrt{M^{2}+ 4\,m^{2}}\simeq M
\label{019}
\ee

The mixing angle $\theta$ is given by
\be
\tan 2\,\theta = \frac{2\,m}{M}\ll 1
\label{020}
\ee

Thus, the see-saw mechanism is based on the assumption
that in addition to the standard Higgs mechanism of the generation of the
Dirac mass term there exist a beyond the SM mechanism of the generation
of the right-handed Majorana mass term, which 
change the lepton number by two and is characterised by
a mass
$M\gg m $. \footnote{It is obvious that for charged particles 
such mechanism does not exist.}
The Dirac mass term mixes left-handed field $\nu_{L}$, the component of doublet, 
and
right- handed singlet field $\nu_{R}$. As a result of this mixing neutrino
acquires small Majorana mass.

In the general case of three generation for neutrino masses we have

\begin{equation}
m_{i} \simeq \frac { (m_{i}^{f})^{2}} {\rm{M}_{i}} \ll m_{i}^{f}\,.
\label{021}
\end{equation}
Here $m_{i}^{f}$ is the mass of quark or lepton in $i$-th family.

Let us stress that if  neutrino masses are of the see-saw origin in this case:
\begin{itemize}
\item
Neutrino with definite masses are Majorana particles.
\item
There are three light neutrinos.
\item
Neutrino masses satisfy the hierarchy
$ m_{1}\ll m_{2}\ll m_{3}\,.$

\item

The heavy Majorana particles must exist.

\end{itemize}

The existence of the heavy Majorana particles, see-saw partners of neutrinos,
could be a source of the barion asymmetry of the Universe (see \cite{Buch}).

\begin{center}
\section{Neutrino oscillations}
\end{center}

In the case of neutrino mixing it is important to distinguish
flavour neutrinos $\nu_{e}$, $\nu_{\mu}$ and $\nu_{\tau}$ and neutrinos with
definite masses $\nu_{1}$, $\nu_{2}$,...
The flavour neutrinos 
are particles that take part in the standard weak interaction.
For example, neutrino that is produced together with $\mu^{+}$
in the decay
 $\pi^{+}\to \mu^{+} + \nu_{\mu}$ 
is
the muon neutrino $\nu_{\mu}$,
electron antineutrino $\bar \nu_{e}$ produces
$e^{+}$ in the process $\bar \nu_{e}+ p \to e^{+}+n $ 
etc. 

In order to determine {\em the states of the flavour neutrinos}
let us consider a decay
\be
a \to b + l^{+}+ \nu_{l}\,,
\label{022}
\ee
If there is neutrino mixing 
$$
\nu_{\alpha L} = \sum_{i} U_{\alpha i} \nu_{iL}
$$
the state of the final particles is given
\be
 |f> = \,~~\sum_{i}|b\rangle \,~ |l^{+}\rangle \,~ 
|\nu_i\rangle \,~ \langle i\,l^{+}\,b\,| S |\,a\rangle \,,
\label{023}
\ee
where $|\nu_i\rangle$ is the state of neutrino with momentum
$\vec p$ and energy 
$$E_i = \sqrt{p^2 + m_i^2 } \simeq p + \frac{ m_i^2 }{ 2 p }; \,~
( p^2 \gg m_i^2 )$$
and $\langle i\,l^{+}\,b\,| S |\,a\rangle $ 
is the element of S-matrix.

We will assume that the mass- squared differences 
$\Delta m^{2}_{ik}= m^{2}_{i}- m^{2}_{k}$ 
are so small that emission of neutrinos with 
different masses can not be resolved in the neutrino production
(and detection) experiments.
 In this case we have

\be
\langle i\,l^{+}\,b\,| S |\,a\rangle \simeq U_{li}^{*}\,~
\langle \nu_{l}\,l^{+}\,b\,| S |\,a\rangle_{SM}\,,
\label{024}
\ee
where 
$ \langle \nu_{l}\,l^{+}\,b\,| S |\,a\rangle_ {SM}$
is the Standard Model matrix element of the process (\ref{022}).

From (\ref{023}) and (\ref{024}) for the normalised state of the flavour 
neutrino $\nu_l$ we obtain

\begin{equation}
|\nu_l\rangle
=
\sum_{i} U_{li}^* \,~ |\nu_i\rangle
\,.
\label{025}
\end{equation}

Thus, in the case of the mixing of the fields of neutrinos with
small neutrino mass-squared differences the state of flavour neutrino
is {\em a coherent superposition} of the states of neutrinos with definite masses.\footnote{The relation (\ref{025}) is analogous to the relations that connects
the states of $K^{0}$ and $\bar K^{0}$ mesons, particles with definite strangeness, with the states of $K_{S}^{0}$ and $ K_{L}^{0}$ mesons, particles with
definite masses and widths.}

In the general case of active and sterile neutrinos 
we have

\be
|\nu_{\alpha}\rangle
= \sum_{i} U_{\alpha i}^* \,~ |\nu_i\rangle \,,
\label{026}
\ee
where index $\alpha $ takes the values $e, \mu, \tau, s_{1},...$. 
From the unitarity of the mixing matrix it follows that 
\be
\langle\nu_{\alpha'}|\nu_{\alpha}\rangle = \delta_{{\alpha'}\alpha}.
\label{027}
\ee

The phenomenon of neutrino oscillations is based on the relation (\ref{026}).
Let us consider the evolution of the mixed neutrino states in vacuum.
If at the initial time $t=0$ flavour neutrino $\nu_{\alpha}$
 is produced, for the neutrino state at the time $t$ we have

\be
|\nu_{\alpha}\rangle_{t}=\,~ e^{-iH_{0}\,t}\,~|\nu_{\alpha}\rangle =
\sum_{i}\,U_{\alpha i}^*\,e^{-iE_it}\,|i\rangle.
\label{028}
\ee

Because of different neutrino masses, phase factors in 
(\ref{028}) are different. This means that the  flavour content of the
final state 
differs from the initial one. At macroscopic distances this effect can be large in spite of small differences of neutrino masses.

Neutrinos are detected through the observation of CC and NC processes.
Developing the state $|\nu_{\alpha}\rangle_{t}$ over the total system of
the flavour (and sterile) neutrino states $|\nu_{\alpha}\rangle$, we have

\be
|\nu_{\alpha}\rangle_{t} =\sum_{\alpha'}A(\nu_\alpha \to \nu_{\alpha'})\,
|\nu_{\alpha'}\rangle\,,
\label{029}
\ee
where
\be
A(\nu_\alpha \to \nu_{\alpha'}) = 
 \langle \nu_{\alpha'}\,|e^{-iH_{0}\,t}\,|\nu_{\alpha}\rangle =
 \sum_i U_{\alpha' i}\,~e^{-iE_it}\,~U_{\alpha i}^*\,.
\label{030}
\ee
is
the amplitude of the transition $\nu_{\alpha} \to \nu_{\alpha'}$ during the time $t$.

Taking into account
the unitarity of the mixing matrix, for
the probability of the transition $\nu_{\alpha} \to \nu_{\alpha'}$
we obtain the following expression \footnote{ We label neutrino masses in such a way that $m_1 < m_2 < m_3<...$}

\begin{equation}
{\mathrm P}(\nu_\alpha \to \nu_{\alpha'}) =
|\delta_{{\alpha'}\alpha} +\sum_{i} U_{\alpha' i}  U_{\alpha i}^*
\,~ (e^{- i \Delta m^2_{i 1} \frac {L} {2E}} -1)|^2 \,,
\label{031}
\end{equation}
where $L\simeq t$ is the distance between neutrino source and neutrino 
detector and $E$
is the neutrino energy. 

Analogously, for the probability of the transition
$\bar\nu_{\alpha} \to \bar \nu_{\alpha'}$ we have

\begin{equation}
{\mathrm P}(\bar\nu_\alpha \to \bar\nu_{\alpha'}) =
|\delta_{{\alpha'}\alpha} +\sum_{i} U_{\alpha' i}^*  U_{\alpha i}
\,~ (e^{- i \Delta m^2_{i 1} \frac {L} {2E}} -1)|^2 \,.
\label{032}
\end{equation}

Let us notice the following general features of the transition
probabilities (see, for example, \cite{BPet,BGG}):

\begin{itemize}

\item Transition probabilities depend on $\frac{L}{E}$.

\item Neutrino oscillations can be observed if
the condition
$\Delta m^2_{i 1} \frac {L} {E}\gtrsim 1 $
is satisfied for at least one value of $ i$.

\item From the comparison of (\ref{031}) and (\ref{032})
we conclude that the following relation holds

$${\mathrm P}(\nu_\alpha \to \nu_{\alpha'}) =
{\mathrm P}(\bar \nu_{\alpha'}  \to \bar\nu_{\alpha}). $$
This relation is the consequence of the CPT invariance
intrinsic for any local field theory.

\item
In the case of the CP invariance in the lepton sector the mixing matrix
$U$ is real in the Dirac case. In the Majorana case 
the mixing matrix 
satisfies the condition
\be
U_{\alpha i}=U_{\alpha i}^{*}\,~\eta_{i}\,,
\label{033}
\ee
where $\eta_{i}=\pm i$ is the CP parity of the Majorana neutrino $\nu_{i}$.
From (\ref{031}), (\ref{032}) and (\ref{033})
we conclude that in the case of the CP invariance in the lepton sector
we have the following relation
$${\mathrm P}(\nu_\alpha \to \nu_{\alpha'}) =
{\mathrm P}(\bar\nu_\alpha \to \bar\nu_{\alpha'}). $$
\end{itemize}

\section{Oscillations between two types of neutrinos}

We will consider here the simplest case of the transitions between two types of neutrinos ($\nu_{\mu} \to \nu_{\tau}$ or  $\nu_{\mu} \to \nu_{e}$ etc). 
In this case the index $i$ in Eq.(\ref{031}) 
takes only one value $i =2$ and 
for the transition probability we obtain expression
\be
{\mathrm P}(\nu_\alpha \to \nu_{\alpha'}) =
|\delta_{{\alpha'}\alpha} + U_{\alpha' 2}  U_{\alpha 2}^*
\,~ (e^{- i \Delta m^2 \frac {L} {2E}} -1)|^2 \,,
\label{034}
\ee
where $\Delta m^2= m^2_{2}-m^2_{1} $.

From this expression for the appearance probability 
($\alpha' \not= \alpha$) we have
\begin{equation}
{\mathrm P}(\nu_\alpha \to \nu_{\alpha'}) =
= \frac {1} {2} {\mathrm A}_{{\alpha'};\alpha}\,~
 (1 - \cos \Delta m^{2} \frac {L} {2E})\,,
\label{035}
\end{equation}
where the amplitude ${\mathrm A}_{{\alpha'};\alpha}$ is given by 
$${\mathrm A}_{{\alpha'};\alpha}= 4\,~|U_{\alpha' 2}|^{2}
\,~|U_{\alpha 2}|^{2} = {\mathrm A}_{{\alpha};\alpha'}\,.$$

Let us introduce the mixing angle $\theta$.
We have 
$$|U_{\alpha 2}|^{2} = \sin^{2}\theta;\,~
|U_{\alpha' 2}|^{2}=1-|U_{\alpha 2}|^{2} = \cos^{2}\theta $$
 and the amplitude
${\mathrm A}_{{\alpha'};\alpha}$
is given by
$${\mathrm A}_{{\alpha'};\alpha}
=\sin^{2}  2\theta \,.$$

Hence, the two-neutrino
transition probability takes the standard form
\begin{equation}
{\mathrm P}(\nu_\alpha \to \nu_{\alpha'}) =
=\frac {1}{2}\,~  \sin^{2}  2\theta \,~
 (1 - \cos \Delta m^{2} \frac {L} {2E})\,.
\label{036}
\end{equation}

It is obvious that in the two-neutrino case the following relations are
valid 
\be
{\mathrm P}(\nu_{\alpha} \to \nu_{\alpha'})=  
{\mathrm P}(\nu_{\alpha'}  \to \nu_{\alpha}) =
{\mathrm P}(\bar\nu_{\alpha} \to \bar\nu_{\alpha'});\,~~(\alpha' \not= \alpha).
\label{037}
\ee
Thus, the CP violation in the lepton sector can not be revealed in the case
of the transitions between two types of neutrinos.

The survival probability ${\mathrm P}(\nu_\alpha \to \nu_{\alpha})$
is determined by condition of the conservation of the probability. We have 
\be
{\mathrm P}(\nu_\alpha \to \nu_\alpha) 
=1-{\mathrm P}(\nu_\alpha \to \nu_{\alpha'})=
 1 - \frac {1}{2}\,~  \sin^{2}  2\theta \,~
(1 - \cos \Delta m^{2} \frac {L}{2E})\,.
\label{038}
\ee
From (\ref{037}) it follows that the two-neutrino survival probabilities
satisfy the following relation
\be
{\mathrm P}(\nu_\alpha \to \nu_{\alpha})=  
{\mathrm P}(\nu_{\alpha'}  \to \nu_{\alpha'})\,.
\label{039}
\ee
Thus, in the case of the transition between two types of neutrinos 
{\em all transition
probabilities} are characterised by the two oscillation parameters:
$\sin^{2} 2\theta$ and  $\Delta m^{2}$.

The expressions (\ref{036})
and (\ref{038}) describe periodical transitions between two types of neutrinos
(neutrino oscillations). They are widely used in the 
analysis of 
experimental data.

The expression (\ref{036}) for the two-neutrino transition
probability 
can be written in the form

\be
{\mathrm P}(\nu_\alpha \to \nu_{\alpha'}) =
= \frac {1} {2} \,\sin^{2}  2\theta\,~
 (1 - \cos 2\,\pi \frac {L} {L_{0}})\,,
\label{040}
\ee
where 
\be
L_{0}=4\,\pi \frac {E} {\Delta m^{2} } 
\label{041}
\ee
is the oscillation length.

Finally, the two- neutrino transition probability and the oscillation length
can be written as
\be
{\mathrm P}(\nu_\alpha \to \nu_{\alpha'})
= \frac {1} {2}\,\sin^{2}  2\theta \,~
 (1 - \cos 2.53 \,\Delta m^{2} \frac {L} {E})
\label{042}
\ee
and
\be
L_{0}\simeq 2.48 \,~\frac {E} {\Delta m^{2} }\,~\rm{m} \,,
\label{043}
\ee
where $E$ is the neutrino energy in MeV, $L$ is the distance in m and 
$\Delta m^{2}$ is neutrino mass-squared difference in $\rm{eV}^{2}$.

\begin{center}
\section{Neutrino oscillation data}
\end{center}

\subsection{Evidence in favour of oscillations of atmospheric neutrinos}

Atmospheric neutrinos are produced mainly  in the decays of pions and
muons

\be
\pi \to \mu  + \nu_{\mu};\,~~
\mu  \to e + \nu_{\mu} + \nu_{e}\,. 
\label{044}
\ee
In the Super-Kamiokande (S-K) experiment \cite{S-K} 
neutrinos are detected via the observation of the 
Cherenkov light emitted by 
electrons and muons
in the large water Cherenkov detector (50 kt of $\rm{H_{2}\,O}$).

At energies smaller than about 1 Gev practically
all muons decay in the atmosphere and from (\ref{044}) it follows that 
$R_{\mu/e}\simeq 2,$
where
$R_{\mu/e}$  is the ratio of the numbers of muon and electron events.

At higher energies the ratio 
$R_{\mu/e}$ 
is larger than
two. It can be predicted, however, with an accuracy better than 5 \%.

The ratio $(R_{\mu/e})$, measured in the S-K \cite{S-K} and SOUDAN 2 
\cite{Soudan} atmospheric neutrino experiments, 
is significantly smaller than the predicted ratio $(R_{\mu/e})_{\rm{MC}}$.
In the S-K experiment for the ratio of ratios 
in the Sub-GeV ($E_{vis}\leq 1.33\,\rm{GeV}$) and Multi-GeV 
 region ($E_{vis}> 1.33\,\rm{GeV}$) regions it was obtained, respectively 

$$\frac{(R_{\mu/e})_{meas}}{(R_{\mu/e})_{\rm{MC}}} = 0.638 \pm 0.016 \pm 0.050;\,~\frac{(R_{\mu/e})_{meas}}{(R_{\mu/e})_{\rm{MC}}}
= 0.658 \pm 0.030 \pm 0.078$$

The fact that the ratio $(R_{\mu/e})_{meas}$ is significantly
smaller than the predicted ratio was known from the results of the
previous atmospheric neutrino experiments Kamiokande \cite{Kam} and IMB 
\cite{IMB}.
During many years
this ``atmospheric neutrino anomaly''  was considered as an indication in favour of neutrino oscillations.

The compelling evidence in favour of neutrino oscillations
was obtained recently by the S-K collaboration 
\cite{S-K} from the observation 
of the large up-down asymmetry of the atmospheric high energy muon events.

If there are no neutrino oscillations, for the number of the electron (muon) events we have the following relation

\be
N_{l}(\cos\theta_{z})= N_{l}( -\cos \theta_{z})\,~~ (l=e,\mu)\,,
\label{045}
\ee
where $\theta_{z}$ is the zenith angle.

For electron events a good 
agreement with this relation was obtained in the S-K experiment.  
For the Multi-GeV muon events the 
significant 
violation of the relation (\ref{045}) was observed.
For the ratio of the 
total number of the up-going muons $U_{\mu}$ 
($\pi/2 \leq \theta_{z}\leq \pi $) to 
the total number of the down-going muons $D_{\mu}$
($0\leq \theta_{z}\leq \pi/2 $)
it was found the value
$$  \left(\frac{U}{D}\right)_{\mu}= 0.54 \pm 0.04 \pm 0.01.$$
At high energies 
leptons are emitted practically in the  
direction of neutrinos.
Thus, up-going muons are produced by neutrinos which travel distances
from $\simeq 500\,\rm{km}$ to  
$\simeq 13000 \, \rm{km}$ and  
the down-going muons are produced by neutrinos which travel distances from $\simeq 20\, \rm{km}$
to $\simeq 500\, \rm{km}$.
The observation of the up-down asymmetry 
clearly demonstrates the dependence of the number of the muon neutrinos
on the distance which they
travel from the production point in the atmosphere to the detector.

The S-K data \cite{S-K} and data of other atmospheric neutrino experiments
(SOUDAN 2 \cite{Soudan}, MACRO \cite{MACRO} ) are well described, 
if we assume that the two-neutrino oscillations $\nu_{\mu}\to
\nu_{\tau}$ take place. 
From the analysis of the S-K data 
it was found that at 90 \% $\rm{CL}$
neutrino oscillation parameters 
$\Delta m^{2}_{atm}$ and $\sin^{2}2 \theta_{atm}$ 
are in the range

$$1.6\cdot 10^{-3}\leq  \Delta m^{2}_{atm}\leq 3.9 \cdot 10^{-3}\,~\rm{ eV}^{2};\,~
\sin^{2}2 \theta_{atm}>0.92;\,.$$

The best-fit values of the parameters are equal
\be
\Delta m^{2}_{atm}=2.5\cdot 10^{-3}\rm{ eV}^{2};\,~\sin^{2}2 \theta_{atm}=1.0 
\,~(\chi^{2}_{\rm{min}}= 163.2/ 170\,\rm{d.o.f.}) 
\label{046}
\ee

\subsection{Evidence in favour of transitions of solar $\nu_{e}$ into
$\nu_{\mu,\tau}$ }

The energy of the sun is produced in the reactions of the thermonuclear
$\rm{pp}$ and $\rm{CNO}$ cycles in which protons and electrons are
converted into helium and electron neutrinos 

$$ 4p + 2 e^{-} \to ^{4}\rm{He}+ 2\nu_{e}.$$

The most important for the solar neutrino experiments reactions 
are listed 
in the Table I. 

\begin{center} 
 Table I
\end{center}
\begin{center} 
The main sources of the solar neutrinos.
The maximum neutrino energies and SSM BP00 fluxes are also given.
\end{center}

\begin{center}
\begin{tabular}{|ccc|}
\hline
Reaction
&
Neutrino energy
&
SSM BP00 flux
\\
\hline
$p\, p \to d\,  e^{+}\, \nu_{e}$
&
$\leq 0.42\, \rm{MeV}$
&
$5.95\cdot 10^{10}\,cm^{-2}\,s^{-1}$
\\
$e^{-}+ ^{7}\rm{Be}\to \nu_{e}\,^{7}\rm{Li}$ 
&
$ 0.86\, \rm{MeV}$
&
$4.77\cdot 10^{9}\,cm^{-2}\,s^{-1}$
\\
$^{8}\rm{B} \to ^{8}\rm{Be}^{*}\,e^{+}\, \nu_{e}$
&
$\leq 15\, \rm{MeV}$
&
$5.05\cdot 10^{6}\,cm^{-2}\,s^{-1}$
\\
\hline
\end{tabular}
\end{center}

As it is seen from the Table I, the major part of the solar neutrino flux 
constitute
the small energy $\rm{pp}$ neutrinos. 
According to the SSM BP00 \cite{BPin} the medium energy monoenergetic 
$^{7}\rm{Be}$
neutrinos make up about 10 \% of the total flux. The high energy
$^{8}\rm{B}$ neutrinos constitute only about $10^{-2}$ \% of the total flux.
However, in the S-K \cite{S-Ksol} and SNO \cite{SNO,SNONC,SNOCC} experiments due to high energy thresholds practically only neutrinos from $^{8}\rm{B}$-decay 
can be detected.\footnote{ According to the SSM BP00 the flux of the high energy $hep$
neutrinos,
produced in the reaction $ ^{3}\rm{He}+ p \to ^{4}\rm{He} + e^{+} + \nu_{e}$,
is about three order of magnitude smaller than the flux of the $^{8}\rm{B}$ neutrinos.}
 $^{8}\rm{B}$ neutrinos give dominant contribution to the event
rate measured in the Homestake experiment \cite{Cl}
experiment. 

The event rates measured in 
all solar neutrino experiments 
are significantly smaller than the event rates, predicted by the Standard Solar
 models.
For the ratio R of the observed and the 
predicted by SSM BP00 \cite{BPin} rates
in the Homestake \cite{Cl}, 
GALLEX-GNO \cite{GALLEX-GNO},
 SAGE \cite{SAGE} and S-K \cite{S-Ksol}  
experiments the following values were obtained:
\bea
 &&R = 0.34 \pm 0.03 \,~~~~ (\mathrm{Homestake})
\nonumber\\
&&R = 0.58 \pm 0.05 \,~~~~ (\mathrm{GALLEX-GNO})
\nonumber\\
 &&R= 0.60 \pm 0.05 \,~~~~ (\mathrm{SAGE})
\nonumber\\
 &&R = 0.465 \pm 0.018  \,~~ (\mathrm{S-K})
\nonumber
\eea

If there is neutrino mixing, original solar $\nu_{e}$'s
due to neutrino oscillations or matter MSW transitions are transfered into another types of neutrinos, which can not be detected by the radiochemical
Homestake, GALLEX-GNO and SAGE experiments. In the S-K experiment
mainly  $\nu_{e}$ are detected: the sensitivity of the
experiment to $\nu_{\mu}$ and $\nu_{\tau}$ is about six times smaller than the
sensitivity to $\nu_{e}$.
Thus, neutrino oscillations or MSW transition in matter
provide the natural explanations of depletion of the fluxes of solar $\nu_{e}$.

Recently 
strong model independent evidence in favour of the transition of the solar
$\nu_{e}$ into  $\nu_{\mu}$ and $\nu_{\tau}$ was obtained in the
SNO experiment \cite{SNO,SNOCC,SNONC}.  
The detector in the SNO experiment  is a 
heavy water Cherenkov detector 
(1 kton of $\rm{D}_{2}O$).
Neutrinos from the sun are detected via the observation of
the following three reactions:\footnote{$\nu_x$ stand for {\em any} flavour neutrino}
\begin{enumerate}

\item 
CC reaction
\be
\nu_e + d \to e^{-}+ p +p\,,
\label{047}
\ee

\item 
NC reaction 
\be
\nu_x + d \to \nu_x + n +p\,,
\label{048}
\ee
\item 
ES process
\be
\nu_x + e \to \nu_x + e \,
\label{049}
\ee
\end{enumerate}

During 306.4 days of running  $1967^{+61.9}_{-60.9}$  
CC events, $576.5^{+49.5}_{-48.9}$ NC events,
and  $263.6^{+26.4}_{-25.6}$ 
ES events were recorded in the SNO experiment. 
The kinetic energy threshold for the detection of electrons was equal to
5 MeV. The NC threshold is 2.2 MeV. 
Thus,
practically only neutrinos from $^{8}\rm{B}$-decay are detected in the SNO experiment.
The initial spectrum of 
electron neutrinos from the decay
$^{8}\rm{B}\to ^{8}\rm{Be}+e^{+}+\nu_{e} $ is known \cite{Ortiz}.

The total CC event rate is given by

\be
R_{\nu_{e}}^{CC}=  <\sigma^{CC}_{\nu_{e}d}>\Phi_{\nu_{e}}^{CC}\,,
\label{050}
\ee
where $<\sigma^{CC}_{\nu_{e}d}>$ is cross section of the process
(\ref{047}),
averaged over known initial spectrum of
$^{8}\rm{B}$ neutrinos, and $\Phi_{\nu_{e}}^{CC}$
is the flux of $\nu_e$ on the earth
The flux $\Phi_{\nu_{e}}^{CC}$ is given by
the relation
\be
\Phi_{\nu_{e}}^{CC} = <P(\nu_e \to\nu_e)>_{CC}\,~\Phi_{\nu_{e}}^{0}\,, 
\label{051}
\ee
where $\Phi_{\nu_{e}}^{0}$ is the total (unknown) initial flux of  $\nu_e$
and $<P(\nu_e \to\nu_e)>_{CC}$ is the averaged $\nu_e$ survival probability.

All flavour neutrinos $\nu_e$, $\nu_{\mu}$
and $\nu_{\tau}$ are recorded via
the detection of the NC process (\ref{048}).
Taking into account $\nu_{e}-\nu_{\mu}- \nu_{\tau}$
universality of the NC for the total NC event rate we have

\be
R^{NC}_{\nu}=  <\sigma^{NC}_{\nu d}>\Phi_{\nu}^{NC}\,,
\label{052}
\ee

where  $<\sigma^{NC}_{\nu d}>$ is the cross section of the process
(\ref{048}), averaged over the initial spectrum of the 
$^{8}\rm{B}$ neutrinos, and  $\Phi_{\nu}^{NC}$
is the total flux of all flavour neutrinos on the earth.
We have 

\be
\Phi_{\nu}^{NC}=\sum_{l=e,\mu,\tau}\Phi_{\nu_l}^{NC}\,.
\label{053}
\ee

Here 
\be
\Phi_{\nu_l}^{NC}=  <P(\nu_e \to\nu_l)>_{NC}\,~\Phi_{\nu_{e}}^{0}\,, 
\label{054}
\ee

where $<P(\nu_e \to\nu_l)>_{NC}$ is the averaged probability of the transition
$\nu_e \to\nu_l$.

All flavour neutrinos are detected also via the observation of the ES process
(\ref{049}). However, the cross section of the (NC) 
$\nu_{\mu,\tau}+e \to\nu_{\mu,\tau}+e$ scattering is about six times smaller
than the cross section of the (CC + NC) $\nu_{e}+e \to\nu_{e}+e$ 
scattering.

The total ES event rate can be presented in the form

\be
R^{ES}_{\nu}=  <\sigma_{\nu_{e} e}>\Phi_{\nu}^{ES}.
\label{055}
\ee

Here $<\sigma_{\nu_{e} e}>$ is the cross section of the process 
$\nu_{e}e \to\nu_{e}e$, 
averaged over initial spectrum of the $^{8}B$ neutrinos and
\be
\Phi_{\nu}^{ES}= \Phi_{\nu_{e}}^{ES} + \frac{<\sigma_{\nu_{\mu} e}>}
{<\sigma_{\nu_{e} e}>}\,~\Phi_{\nu_{\mu,\tau}}^{ES}\,,
\label{056}
\ee
where 
$\Phi_{\nu_{e}}^{ES}$  is 
the flux of $\nu_{e}$, $\Phi_{\nu_{\mu,\tau}}^{ES}$  is the flux of
$\nu_{\mu}$ and $\nu_{\tau}$
and
\be
\frac{<\sigma_{\nu_{\mu} e}>}
{<\sigma_{\nu_{e} e}>}
\simeq 0.154.
\label{057}
\ee

We have
\be
\Phi_{\nu_{l}}^{ES}= <P(\nu_e \to\nu_l)>_{ES}\,~\Phi_{\nu_{e}}^{0},
\label{058}
\ee
where
$<P(\nu_e \to\nu_l)>_{ES}$ is the averaged probability of the transition $\nu_e \to\nu_l$.

In the SNO experiment it was obtained \cite{SNOCC}
\be
(\Phi_{\nu}^{ES})_{\rm{SNO}} =({2.39^{+0.24}_{-0.23}~ \mbox{(stat.)}
\pm 0.12~\mbox{(syst.)}} ) \cdot 10^{6}\,~
cm^{-2}s^{-1} \,,
\label{059}
\ee
This value is in a good agreement with 
the S-K value.
In the S-K experiment \cite{S-Ksol} solar neutrinos are detected via the observation of the
ES process
$\nu_x e\to \nu_x e$. During 1496 days of running a large number $22400 \pm 800$
solar neutrino events 
with recoil total energy threshold 5 MeV
were recorded. From the data of the 
S-K experiment it was obtained 
\be
(\Phi_{\nu_e}^{ES})_{\rm{S-K}} =({2.35 \pm 0.02~ \mbox{(stat.)} \pm0.08~\mbox{(syst.)}} ) \cdot 10^{6}\,~
cm^{-2}s^{-1} \,.
\label{060}
\ee

In the S-K experiment the spectrum of the recoil electrons
was measured. No sizable distortion of the spectrum with respect to the
expected spectrum was observed.
The spectrum of electrons, produced in the CC process (\ref{047}),
was measured in the SNO experiment \cite{SNOCC}. No distortion of the 
electron spectrum was observed also in this experiment.\footnote{Expected spectra were calculated under the assumption that the shape of the
spectrum of $\nu_{e}$ on the earth is given by known initial $^{8} B$ spectrum}

Thus, the data of the S-K and SNO experiments are compatible with the assumption that in the
high-energy $^{8} B$ region the probability of the solar neutrinos to survive is a constant:
\be
P(\nu_e \to\nu_e) \simeq \rm{const}.
\label{061}
\ee

From (\ref{061})
it follows that
$$<P(\nu_e \to\nu_e)>_{CC}\simeq <P(\nu_e \to\nu_e)>_{NC}
\simeq <P(\nu_e \to\nu_e)>_{ES}\,.$$
Taking into account these relations, 
from (\ref{051}), (\ref{054}) and (\ref{058}) 
it follows that in the 
high energy $^{8} B$ region the fluxes of 
electron neutrinos, detected
via the observation of CC, NC and ES processes, are the same:
\be
\Phi_{\nu_{e}}^{CC}\simeq \Phi_{\nu_{e}}^{NC}\simeq
\Phi_{\nu_{e}}^{ES}\,.
\label{062}
\ee

From the data of the SNO experiment \cite{SNONC,SNOCC}
it was obtained that
the flux of $\nu_e$ on the earth
is equal to

\be
(\Phi_{\nu}^{CC})_{\rm{SNO}} =({1.76^{+0.06}_{-0.05}\mbox{(stat.)}^{+0.09}_{-0.09}~\mbox{(syst.)}} ) \cdot 10^{6}\,~
cm^{-2}s^{-1} \,.
\label{063}
\ee

For the flux of all flavour neutrinos $\Phi_{\nu}^{NC}$
it was found the value
\be
(\Phi_{\nu}^{NC})_{\rm{SNO}} =({5.09^{+0.44}_{-0.43}\mbox{(stat.)}^{+0.46}_{-0.43}~\mbox{(syst.)}} ) \cdot 10^{6}\,~
cm^{-2}s^{-1}\,, 
\label{064}
\ee
which is about three times larger than the value 
 of the flux of electron neutrinos.

It is obvious that the NC flux $\Phi_{\nu}^{NC}$
is given by
\be
\Phi_{\nu}^{NC}= \Phi_{\nu_e}^{NC}+ \Phi_{\nu_{\mu,\tau}}^{NC}\,
\label{065}
\ee
where $\Phi_{\nu_e}^{NC}$ is the flux of $\nu_{e}$ and
$\Phi_{\nu_{\mu,\tau}}^{NC}$ is the flux of $\nu_{\mu}$ and 
$\nu_{\tau}$.

Combining CC and NC fluxes and using the relation (\ref{062}), we can determine now 
the flux $\Phi_{\nu_{\mu,\tau}}^{NC}$.
Taking into account also the value (\ref{059})
 of the ES flux, in \cite{SNONC}
it was obtained

\be
(\Phi_{\nu_{\mu,\tau}})_{SNO} =(3.41^{+0.45}_{-0.45}\mbox{(stat.)}^{+0.48}_{-0.45}~\mbox{(syst.)})\cdot 10^{6}\,~
cm^{-2}s^{-1} \,.
\label{066}
\ee

Thus, detection of the solar neutrinos  
via the  simultaneous observation of 
CC, NC and ES processes allowed the SNO collaboration
to obtain {\em the direct model independent $5.3\,~ \sigma$ evidence 
of the presence of
$\nu_{\mu}$ and $\nu_{\tau}$ in the flux of the solar neutrinos on the earth.}

The total flux of the $^{8} B$ neutrinos, predicted by SSM BP00 \cite{BPin},
is given
\be
(\Phi_{\nu_{e}}^{0})_{\rm{SSM\, BP}}= (5.05^{+1.01}_{-0.81})\cdot 10^{6}\,~
cm^{-2}s^{-1}  
\label{067}
\ee

This flux is compatible with the total flux of all flavour neutrinos
(\ref{064}), measured in the SNO experiment.

The flux of $\nu_{\mu}$ and $\nu_{\tau}$ on the earth can be also obtained
from the SNO CC data and the S-K ES data. In first SNO publication 
\cite{SNO} it was found the value

\be
(\Phi_{\nu_{\mu,\tau}})_{\rm{S-K, SNO}} =(3.69 \pm 1.13 )\cdot 10^{6}\,~
cm^{-2}s^{-1} \,,
\label{068}
\ee
which is in a good agreement with the value (\ref{066}).

The data of all solar neutrino experiments can be described if we 
assume that there are transitions of the solar $\nu_e$ into 
$\nu_{\mu,\tau}$ and $\nu_e$ survival probability has two-neutrino form,
which is charachterized by two oscillation parameters  
$\Delta m^{2}_{\rm{sol}}$ and $\tan^{2}\theta_{\rm{sol}}$.
From the global $\chi^{2}$ fit of the total event rates measured in
all solar neutrino experiments several allowed regions in the plane of
the oscillation parameters were obtained (see, for example, \cite{BGP}): large mixing angle MSW
LMA and LOW regions, small mixing angle MSW SMA region,
vacuum oscillations VO region and others. The situation changed after
the day and night recoil electron spectra were measured in the S-K experiment 
\cite{S-Ksol} and  SNO data \cite{SNO,SNOCC,SNONC} were obtained. 
From all analysis of the existing 
solar neutrino data it follows that the most plausible 
allowed region is the MSW LMA region (see \cite{Lisi} and references therein).

In \cite{SNOCC} as a free variable parameters 
$\Delta m^{2}_{\rm{sol}}$ , $\tan^{2}\theta_{\rm{sol}}$ and the initial
flux of the $^{8} B$ neutrinos  $\Phi_{\nu_{e}}^{0}$ were used . From the analysis of all
solar neutrino data the following best-fit values of the parameters 
were found ($\chi^{2}_{\rm{min}}= 57/72\, \rm{d.o.f.}$):

\be
\Delta m^{2}_{\rm{sol}}=5\cdot 10^{-5}\rm{eV}^{2};
\,\tan^{2}\theta_{\rm{sol}}=0.34\,~
\Phi_{\nu_{e}}^{0}=5.89\cdot 10^{6}\,cm^{-2}\,s^{-1}
\label{069}
\ee

If neutrino oscillation parameters 
are in the LMA region, neutrino oscillations
in the solar range of $\Delta m^{2}$
can be explored in experiments with reactor $\bar\nu_{e}$'s
if a distance between reactors and a detector is about 100 km.
In the experiment KamLAND \cite{KamL}, which started in January
2002,  $\bar\nu_{e}$ 
from several Japanese reactors are recorded by a large liquid 
scintillator detector
(1 kt of liquid 
scintillator).
The distance between reactors and the detector is $ 175 \pm 35 $ km.
The average energy of $\bar\nu_{e}$ from a reactor is about 3 MeV.
Thus, at large mixing angles the KamLAND experiment
is sensitive to the solar LMA range of neutrino mass-squared difference
($\Delta m^{2}\simeq \frac{E}{L}\simeq 10^{-5}\rm{eV}^{2}$).

\subsection{Reactor experiments CHOOZ and Palo Verde}

The results of the long baseline 
reactor experiments CHOOZ \cite{CHOOZ} and Palo Verde \cite{PaloV}
 are very
important for the neutrino mixing.
In these experiments the disappearance of the reactor $\bar\nu_{e}$'s
in the atmospheric range of $\Delta m^{2}$ were searched for.

In the CHOOZ experiment $\bar\nu_{e}$ from two reactors at the distance of
about 1km from the detector were detected via the observation of the process
$$\bar\nu_{e}+ p \to e^{+}+ n. $$ 
No indications in favour of disappearance of $\bar\nu_{e}$ were found
in the experiment. For the ratio $R$ of the total number of the
detected $\bar\nu_{e}$ events
to the expected number it was found the value

$$ R =1.01 \pm 2.8 \%\,(\rm{stat})\pm\pm 2.7 \%\,(\rm{syst})\,~~( \rm{CHOOZ})$$
In the similar Palo Verde experiment it was found:

$$R =1.01 \pm 2.4 \%\,(\rm{stat})\pm 5.3 \%\,(\rm{syst})
\,~~(\rm{Palo Verde}) $$

The data of the experiments were analysed in \cite{CHOOZ,PaloV} in the framework of 
two-neutrino oscillations and exclusion plots in the plane of the oscillation parameters $\Delta m^{2}$ and $\sin^{2}2\,\theta$ were obtained.
From the CHOOZ exclusion plot at $\Delta m^{2}=2.5 \cdot 10^{-3}
\,~\rm{eV}^{2}$ (the S-K best-fit value) we have
$$\sin^{2}2\,\theta \lesssim 1.5\cdot 10^{-1}.$$

\begin{center}
\section{Neutrino oscillations in the framework of three-neutrino mixing}
\end{center}
\subsection{Neutrino oscillations in the atmospheric range of $\Delta m^{2}$}

We have discussed evidences in favour of neutrino oscillations that
 were obtained
in the solar and atmospheric neutrino experiments.
There exist at present also an indication in favour of the transitions
$\bar \nu_{\mu} \to \bar \nu_{e}$, that was obtained in the single accelerator
experiment LSND \cite{LSND}. The LSND data can be explained
by neutrino oscillations. From analysis of the data for the 
values of the oscillation parameters it was obtained the ranges
$$ 2\cdot 10^{-1} \lesssim \Delta m^{2}\lesssim 1\,\rm{eV}^{2};\,~
3\cdot 10^{-3} \lesssim \sin^{2}2\,\theta \lesssim 4\cdot 10^{-2}$$

In order to describe
the data of the solar, atmospheric and LSND experiments, which requires three
different values of neutrino mass-squared differences, it is necessary to assume mixing of (at least) four massive neutrinos (see, for example, \cite{BGG}).

The result of the LSND experiment requires, however, confirmation. 
MiniBooNE experiment at Fermilab \cite{MiniB}, that started in 2002,
is aimed to check the LSND result.

We will consider here the {\em minimal}
scheme of three neutrino mixing

\be
\nu_{\alpha L}=\sum_{i=1}^{3}\,U_{\alpha i}\,\nu_{i L},
\label{070}
\ee
which 
provide two independent $\Delta m^{2}$ and allow
to describe solar and atmospheric neutrino oscillation data.
In (\ref{070}) $U$ is the unitary 3$\times$3 PMNS mixing matrix
\cite{BPon,MNS}.

Let us consider first neutrino oscillations in the atmospheric range
of $\Delta m^{2}$, which can be explored in the atmospheric and long 
baseline accelerator and reactor neutrino experiments.
In the framework of the three-neutrino mixing 
with $m_1 < m_2< m_3$ there are two possibilities:

I. Hierarchy of neutrino mass-squared differences
\be
\Delta m^{2}_{21}\simeq \Delta m^{2}_{\rm{sol}};
\,
\Delta m^{2}_{32}\simeq \Delta m^{2}_{\rm{atm}};
\,~\Delta m^{2}_{21} \ll \Delta m^{2}_{32}.
\label{071}
\ee

II. Inverted hierarchy of neutrino mass-squared differences
\be
\Delta m^{2}_{32}\simeq \Delta m^{2}_{\rm{sol}};
\,
\Delta m^{2}_{21}\simeq \Delta m^{2}_{\rm{atm}};
\,~\Delta m^{2}_{32} \ll \Delta m^{2}_{21}.
\label{072}
\ee

We will assume that neutrino mass spectrum is of the type I.
For the values $\frac{L}{E}$ relevant for neutrino oscillations in the atmospheric range
of $\Delta m^{2}$ ($\Delta m^{2}_{32}\,\frac{L}{E}\gtrsim 1$) we have

$$\Delta m^{2}_{21}\,~ \frac{L}{E}\ll 1.$$
Hence we can neglect 
the contribution of
$\Delta m^{2}_{21}$ to 
the transition probability Eq.(\ref{031}).
For the
probability of the transition
$\nu_\alpha \to \nu_{\alpha'}$
we obtain in this case the following expression

\be
{\mathrm P}(\nu_\alpha \to \nu_{\alpha'}) \simeq
|\delta_{{\alpha'}\alpha} + U_{\alpha' 3}  U_{\alpha 3}^*
\,~ (e^{- i \Delta m^2_{32} \frac {L} {2E}} -1)|^2 
\label{073}
\ee

Thus, in the leading approximation  
transition probabilities
in the atmospheric range of $\Delta m^2$ are determined by the largest neutrino mass-squared difference
$\Delta m^2_{32}$ and the elements  of the third column 
of the neutrino mixing matrix, which connect
flavour neutrino fields $\nu_{\alpha L}$
 with the field of the heaviest neutrino $\nu_{3L}$.

For the appearance probability 
from (\ref{073}) we obtain
\begin{equation}
{\mathrm P}(\nu_\alpha \to \nu_{\alpha'}) =
 \frac {1} {2} {\mathrm A}_{{\alpha'};\alpha}\,~
 (1 - \cos \Delta m^{2}_{32} \frac {L} {2E})\,~(\alpha \not= \alpha'),
\label{074}
\end{equation}

where the oscillation amplitude is given by the expression
\be
{\mathrm A}_{{\alpha'};\alpha}= 4\,~|U_{\alpha' 3}|^{2}\,~|U_{\alpha 3}|^{2}
\label{075}
\ee

The survival probability can be obtained from 
the condition of the conservation of probability and Eq. (\ref{074}).
We have
\be
{\mathrm P}(\nu_\alpha \to \nu_\alpha)= 1 -\sum_{\alpha' \not= \alpha}\,
 {\mathrm P}(\nu_\alpha \to \nu_{\alpha'}) =1 - \frac {1} {2}
{\mathrm B}_{\alpha ; \alpha}\,~ 
(1 - \cos \Delta m^{2}_{32} \frac {L} {2E})\,.
\label{076}
\ee
Taking into account the unitarity of the mixing matrix, for
the oscillation amplitude ${\mathrm B}_{\alpha ; \alpha}$ we have 
\be
{\mathrm B}_{\alpha ; \alpha}=\sum_{\alpha' \not= \alpha}\,
{\mathrm A}_{{\alpha'};\alpha}=
4\,~|U_{\alpha 3}|^{2}
\,~(1 -|U_{\alpha 3}|^{2}).
\label{077}
\ee
Let us notice that in the case of the inverted hierarchy of the 
neutrino mass squared differences
 transition probabilities 
can be obtained from (\ref{074})-(\ref{077})
by the change 
$\Delta m^{2}_{32}\to \Delta m^{2}_{21}$ and $|U_{\alpha 3}|^{2}\to
|U_{\alpha 1}|^{2}$.

Transition probabilities Eq.(\ref{074}) and Eq.(\ref{075}) depend only on 
$|U_{\alpha 3}|^{2}$ and $\Delta m^{2}_{32}$. The CP phase does not
enter into expressions for the transition probabilities. 
This means that in the leading approximation the 
relation

\be
{\mathrm P}(\nu_\alpha \to \nu_{\alpha'})=
{\mathrm P}(\bar \nu_\alpha \to \bar \nu_{\alpha'})
\label{078}
\ee
is satisfied.

Thus, 
investigation of effects of the CP violation in the lepton sector
in the long baseline
neutrino oscillation experiments
will be a difficult
problem: possible effects are suppressed due to the smallness of the
parameter $\frac{\Delta m^{2}_{12}}{\Delta m^{2}_{32}}$.
High precision experiments on the search for effects 
of the CP-violation in the lepton
sector are planned for future Neutrino Superbeam facilities 
\cite{Nakaya} and Neutrino Factories \cite{Lindner,Dydak}.

Transition probabilities (\ref{074})-(\ref{077}) 
have {\em two-neutrino form} in every channel.
This is obvious consequence of the fact that only the largest mass-squared difference
$\Delta{m}^2_{32} $ contributes to the transition probabilities.
The elements  $|U_{\alpha 3}|^{2}$, which determine the oscillation amplitudes,
satisfy
the unitarity condition $\sum_{\alpha}|U_{\alpha 3}|^{2}=1 $. Hence, in the leading approximation transition probabilities are characterised by three parameters.
In 
the standard parametrisation
of the neutrino mixing matrix (see \cite{PDG}) we have

\be
U_{\mu 3} = \sqrt{ 1 -|U_{e 3}|^{2}}\,~ \sin\theta_{23};\,
U_{\tau 3} = \sqrt{ 1 -|U_{e 3}|^{2}}\,~ \cos\theta_{23}\,,
\label{079}
\ee
where $\theta_{23}$ is the mixing angle.

From (\ref{075}),
 and (\ref{079})
for the amplitude of the transition $\nu_{\mu}\to\nu_{\tau}$
and $\nu_{\mu}\to\nu_{e}$ we will obtain, respectively

\be
{\mathrm A}_{\tau;\mu}=(1 -|U_{e 3}|^{2})^{2}\,~\sin^{2}2\theta_{23};\,
{\mathrm A}_{e;\mu}=4\,|U_{e 3}|^{2}\,(1 -|U_{e 3}|^{2})\,
\sin^{2}\theta_{23}\,.
\label{080}
\ee

For the amplitude ${\mathrm B}_{e;e}$ we have\footnote{Notice the following relation
between oscillation amplitudes
$${\mathrm A}_{e;\mu}={\mathrm B}_{e;e}\,~\sin^{2}\theta_{23}\,.$$}
\be
{\mathrm B}_{e;e}=4\,|U_{e 3}|^{2}\,~(1 -|U_{e 3}|^{2})\,.
\label{081}
\ee
In the S-K atmospheric neutrino experiment \cite{S-K}
no any indications in favour of $\nu_{\mu}\to \nu_{e}$
transitions were obtained.
The data of the experiment are
well described under the assumption $|U_{e 3}|^{2}\simeq 0$. In this approximation oscillations
in the atmospheric range of $\Delta m^2$ are pure 
$\nu_{\mu}\to \nu_{\tau}$
two- neutrino oscillations. The values of the two-neutrino oscillation parameters
$\Delta m^2_{atm} \simeq\Delta m^2_{32} $
and $\sin^{2}\theta_{atm}=sin^{2}\theta_{23} $, obtained from the analysis of the S-K data,
 are given in (\ref{046}).

\subsection{Oscillations in the solar range of $\Delta m^2$}

Let us consider now in the framework of the tree-neutrino mixing 
neutrino oscillations in the solar range of $\Delta m^{2}$.
The $\nu_{e}$ survival probability in vacuum
can be written in the form

\be
{\mathrm P}(\nu_e\to\nu_e)=
=
\left|
\sum_{i=1, 2}| U_{e i}|^2 \, 
 e^{ - i
 \, 
\Delta{m}^2_{i1} \frac {L}{2 E} }
 + | U_{e 3}|^2  \, 
 e^{ - i
 \, 
\Delta{m}^2_{31} \frac {L}{2 E} }\,\right|^2
\label{082}
\ee

We are interested in the survival probability averaged over
the region where neutrinos are produced, neutrino energy spectrum etc. 
Because $\Delta m^{2}_{32}$ is much larger than $\Delta m^{2}_{21}$, 
in the averaged survival probability
the interference between the first and the second 
terms in (\ref{082})
disappears. The averaged survival probability can be presented in the form

\be
{\mathrm P}(\nu_{e}\to\nu_{e})=|U_{e 3}|^{4}+ (1-|U_{e 3}|^{2})^{2}\,~
P^{(1,2)}(\nu_{e}\to\nu_{e})\,.
\label{083}
\ee

Here $P^{(1,2)}(\nu_{e}\to\nu_{e})$ 
is given by the expression

\begin{equation}
{\mathrm P}^{(1,2)}(\nu_e \to \nu_e) =
 1 - \frac {1} {2}\,~A^{(1,2)}\,~ 
(1 - \cos \Delta m^{2}_{21} \frac {L} {2E})\,,
\label{084}
\end{equation}

where
\be
A^{(1,2)}= 4\,\frac{ |U_{e 1}|^{2}\,|U_{e 2}|^{2}} 
{( 1-   |U_{e 3}|^{2})^{2} }.
\label{085}
\ee

In the standard parametrisation of the neutrino mixing matrix we have

\be
U_{e 1} = \sqrt{ 1 -|U_{e 3}|^{2}}\,~ \cos\theta_{12}\,;
U_{e 2} = \sqrt{ 1 -|U_{e 3}|^{2}}\,~ \sin\theta_{12}\,,
\label{086}
\ee
where $\theta_{12}$
is the mixing angle. From (\ref{085}) and (\ref{086})
for the amplitude $A^{(1,2)}$ we obtain the following expression

\be
A^{(1,2)}=\sin^{2}2\,\theta_{12}
\label{087}
\ee
Thus, the probability ${\mathrm P}^{(1,2)}(\nu_e \to \nu_e)$
is characterised by two parameters and have the standard two-neutrino form.

The expression (\ref{083}) is also valid in the case
of matter \cite{Schramm,BGG}. In this case
$P^{(1,2)}(\nu_{e}\to\nu_{e})$ is the two-neutrino $\nu_{e}$ survival
probability in matter. In the calculation of this quantity the
density of electrons $\rho_{e}(x)$ in the effective
Hamiltonian of the interaction of neutrino with matter must be
changed by $(1-|U_{e 3}|^{2})\,\rho_{e}(x)$.

As we will see in the next subsection, from the data of
the reactor CHOOZ and Palo Verde experiments it follows that element
$|U_{e 3}|^{2}$ is small. If we neglect $|U_{e 3}|^{2}$ in Eq.
(\ref{083}) we come to the conclusion that in the framework of the
three-neutrino mixing $\nu_e$ survival probability in
the solar range of neutrino mass-squared difference has two-neutrino
form

\be
{\mathrm P}(\nu_e \to \nu_e)\simeq {\mathrm P}^{(1,2)}(\nu_e \to \nu_e)
\label{088}
\ee

The values of the parameters $\Delta m^{2}_{\rm{sol}}\simeq \Delta m^{2}_{21}$
and $\tan^{2}\theta_{\rm{sol}}\simeq\tan^{2}\theta_{12}$,
obtained from the analysis of the solar neutrino data, are given in
(\ref{069}).

Thus, 
due to the smallness of the parameter $|U_{e 3}|^{2}$ 
and  hierarchy of neutrino mass squared differences 
$\Delta m^{2}_{12}\ll\Delta m^{2}_{32}$ 
neutrino
oscillations in the atmospheric and solar ranges of $\Delta m^{2}$ 
in the leading approximation are decoupled \cite{BG} and are described by two-neutrino
formulas, which are characterised by the parameters 
$\Delta m^{2}_{32},\,
\sin^{2}2\,\theta_{23}$
and
$\Delta m^{2}_{21},\,
\tan^{2}\theta_{12}$, respectively.

\subsection{The upper bound of $|U_{e 3}|^{2}$ from the data of the CHOOZ experiment}

The reactor long baseline CHOOZ \cite{CHOOZ} and Palo Verde \cite{PaloV}
experiments are sensitive to the atmospheric range of $\Delta m^{2}$. No any indications
in favour of disappearance of reactor $\bar\nu_e$  were obtained in these experiments.
From the analysis of the data of the CHOOZ and Palo Verde experiments the best bound on the parameter
$|U_{e 3}|^{2}$ can be obtained.

In the framework of the three-neutrino mixing the probability of 
 $\bar\nu_e$ to survive is given 
by the expression

\be
{\mathrm P}(\bar \nu_e \to \bar \nu_e) =1 - \frac {1} {2}
{\mathrm B}_{e; e}\,~ 
(1 - \cos \Delta m^{2}_{32} \frac {L} {2E})\,,
\label{089}
\ee
where the amplitude
$ {\mathrm B}_{e ; e}$ is given by Eq. (\ref{081}).

In \cite{CHOOZ,PaloV} exclusion plots in the plane of the parameters
$\Delta m^{2}\equiv \Delta m^{2}_{32}$ and $\sin^{2}2\theta \equiv {\mathrm B}_{e; e}$
were obtained. From these exclusion plots we have
\be
{\mathrm B}_{e ; e} \leq {\mathrm B}_{e ; e}^{0},
\label{090}
\ee
where the upper bound ${\mathrm B}_{e ; e}^{0}$
depends on 
$\Delta m^{2}$.
For the S-K \cite{S-K} allowed values of $\Delta m^{2}_{32}$  
from the CHOOZ exclusion plot we find
\be
1\cdot 10^{-1}\leq {\mathrm B}_{e ; e}^{0}\leq 2.4\cdot 10^{-1}.
\label{091}
\ee

Using (\ref{081}) and (\ref{090}), for the parameter 
$|U_{e 3}|^{2}$
we have the bounds

\be
|U_{e 3}|^{2} \leq
\frac{1}{2}\,\left(1 - \sqrt{1- {\mathrm B}_{e ; e}^{0} }\right)\lesssim
\frac{1}{4}\,  {\mathrm B}_{e ; e}^{0}
\label{092}
\ee

or
\begin{equation}
|U_{e 3}|^{2} \gtrsim
\frac{1}{2}\,\left(1 + \sqrt{1- {\mathrm B}_{e ; e}^{0}}\right)\geq
1-\frac{1}{4}\, {\mathrm B}_{e ; e}^{0}
\label{093}
\end{equation}

Thus, parameter
$|U_{e 3}|^{2}$ can be small or large (close to one).
This last possibility is excluded by the
solar neutrino data. In fact, if  $|U_{e 3}|^{2}$ is large, from 
Eq. (\ref{083}) it follows that in the whole range of the solar neutrino energies
the probability of $\nu_{e}$ to survive is close to one 
in obvious contradiction with the solar neutrino data.
Thus, the upper bound of the parameter $|U_{e 3}|^{2}$ is given by
(\ref{092}). 
At the
S-K best-fit point 
$\Delta m^{2}_{32}=2.5\cdot 10^{-3}\rm{ eV}^{2}$
we have
\be
|U_{e 3}|^{2}\leq 4\cdot 10^{-2}\,~ (95 \%\,~\rm{CL}).
\label{094}
\ee

\section{Conclusion}

Compelling evidences in favour of neutrino oscillations, driven by
small neutrino masses and neutrino mixing, were obtained in recent years
in the S-K \cite{S-K}, SNO \cite{SNO,SNOCC,SNONC} and other atmospheric and solar neutrino experiments.
These findings opened a new field of research in the particle physics and astrophysics: {\em physics of massive and mixed neutrinos}.

From the results of the experiments it follows that
neutrino masses are
many orders of magnitude smaller than 
the masses of other fundamental fermions (leptons and quarks). 
There is a general consensus that tiny neutrino masses are of a beyond the Standard Model origin.

There are many unsolved problems in the physics of massive and mixed neutrinos.
In the nearest years LMA solution of the solar neutrino problem
will be tested by the  KamLAND \cite{KamL} and the BOREXINO 
\cite{BOREXINO} experiments.
If neutrino oscillation parameters $\Delta m^{2}_{\rm{sol}}$
and $\tan^{2}\theta_{\rm{sol}}$ are in the LMA region, 
neutrino oscillations in the solar range of $\Delta m^{2}$
can be studied in details in terrestrial experiments with well known antineutrino spectrum. 

Another problem which will be probably solved in the nearest years is the
problem of LSND \cite{LSND}.
 If LSND result will be confirmed by the MiniBOONE 
experiment \cite{MiniB} it will mean that the number of light neutrinos
is more than three and in addition to the three flavour neutrinos sterile neutrino(s) must
exist. If LSND result will be not confirmed, the minimal scheme with 
three massive and mixed neutrinos
will be 
very plausible possibility.

The problem of {\em the nature of massive neutrinos (Dirac or Majorana?)}
is one of the most fundamental one. This problem can be solved by the
experiments on the search for neutrinoless double $\beta $-
decay ($(\beta \beta)_{0\nu}$ -decay). 
If massive neutrinos are Majorana particles the matrix element of this 
process is proportional to the effective Majorana mass
$<m> = \sum_{i}U^{2}_{ei}\,m_{i}$.  The most stringent lower bounds
for the time of life of the $(\beta \beta)_{0\nu}$ -decay were obtained
in the $^{76}\rm{Ge}$ experiments \cite{H-M,IGEX}. 
Taking into account different calculations of the nuclear matrix elements,
for the effective Majorana mass from the results of these experiments 
the following upper bounds was obtained 

$$|<m>|\leq (0.3-1.3)\,~\rm{eV}.$$ 

Many new experiments on the search for 
$(\beta \beta)_{0\nu}$ -decay
are in preparation at present \cite{Cremonesi}.
In these  experiments the sensitivities 
$$
|<m>|\simeq (1.5\cdot 10^{-2} - 1\cdot 10^{-1})\,~ \rm{eV}$$
is planned to be achieved.
Existing neutrino oscillation data allow to obtain some constraint
on the value of the effective Majorana mass (see, for example \cite{BPP}). 
If the number of massive 
neutrinos is equal to three and there is see-saw inspired neutrino mass
hierarchy $m_{1}\ll m_{2}\ll m_{3}$,  the upper bound of the effective Majorana mass
is presumably lower than the sensitivity 
of the
$(\beta \beta)_{0\nu}$-experiments of the next generation.

One of the very important problem of neutrino mixing is the problem of
$|U_{e3}|^{2}$. In order to see effects of the three-neutrino mixing
in future long baseline neutrino experiments 
and, in particular, effects of the CP-violation in the lepton sector,
it necessary that the parameter $\Delta m_{21}^{2}$ was in LMA region and
the parameter $|U_{e3}|^{2}$ was larger than $10^{-4}-10^{-5}$ (see \cite{Lindner}).
Best limit on $|U_{e3}|^{2}$ was found from the data of the reactor experiment CHOOZ \cite{CHOOZ}.
New information on $|U_{e3}|^{2}$ will be obtained in the nearest years in the MINOS \cite{MINOS}, ICARUS \cite{ICARUS} and JHF 
\cite{Itow} experiments.

It is obvious that in order to reveal the real origin of the 
newly discovered phenomenon of 
small neutrino
masses and neutrino mixing a lot of work have to be done.
We would like to notice that it is not for the first time
that breakthrough to a new physics was connected with neutrinos.
The first Fermi theory of the $\beta$ -decay 
was based on the Pauli hypothesis of neutrino.
The phenomenological V-A theory of the weak interactions started with
Landau, Lee and Young and Salam two-component neutrino theory.
The first evidence for the Glashow, Weinberg and Salam Standard Model
of the electroweak interaction was obtained
on neutrino beam in CERN 
(discovery of the NC).

It is a pleasure for me to acknowledge support of the ``Programa de Profesores Visitantes de IBERDROLA de Ciencia y Tecnologia'' and  
Istituto Nacionale
di Fisica Nucleare, sezione di Torino.


\begin{thebibliography}{99}

\bibitem{S-K} Super-Kamiokande Collaboration, S.~Fukuda {\it et al.,}
Phys. Rev. Lett. {\bf 81}, 1562 (1998);\,
 S.~Fukuda {\it et al.,} Phys. Rev. Lett. {\bf 82}, 2644 (1999);\,
S.~Fukuda {\it et al.,} Phys. Rev. Lett. {\bf 85}, 3999-4003 (2000).

\bibitem{Soudan} Soudan 2 Collaboration, W.W.M.Allison \textit{et al.}, 
Physics Letters {\bf B 449} (1999) 137; 

\bibitem{MACRO} MACRO Collaboration, M.Ambrosio et al.
hep-ex/0106049;\, Phys. Lett. B517 (2001) 59 \,~M. Ambrosio et al.
NATO Advanced Research Workshop on Cosmic Radiations,
Oujda (Morocco), 21-23 March, 2001.


\bibitem{Cl}B. T. Cleveland {\it et al.}, Astrophys. J. {\bf
496} (1998) 505.

\bibitem{GALLEX-GNO} GALLEX Collaboration, W. Hampel 
{\it et al.}, Phys. Lett. {\bf B 447} (1999) 127 ;\,
GNO Collaboration,
M. Altmann {\it et al.}, Phys. Lett. {\bf B 490} (2000) 16 ;\,
Nucl.Phys.Proc.Suppl. {\bf 91} (2001) 44.

\bibitem{SAGE} SAGE Collaboration,
J. N. Abdurashitov {\it et al.},
 Phys. Rev. {\bf C 60} (1999) 055801 ; \,Nucl.Phys.Proc.Suppl. {\bf 110}
(2002) 315;
\bibitem{S-Ksol} Super-Kamiokande Collaboration, S.~Fukuda {\it et al.}, Phys. Rev. Lett.
 {\bf 86} (2001) 5651;\, M.Smy, hep-ex/0208004.

\bibitem{SNO} SNO collaboration, Q.R. Ahmad {\it et al.}, Phys. Rev. Lett. 
 {\bf 87}, 071301 (2001).
\bibitem{SNONC}SNO collaboration,
Q.R. Ahmad {\it et al.}, Phys.Rev.Lett. 
{\bf 89}, 011301 (2002); nucl-ex/0204008. 
\bibitem{SNOCC}SNO collaboration, Q.R. Ahmad {\it et al.,} 
Phys.Rev.Lett 
{\bf 89}, 011302 (2002); nucl-ex/0204009.







\bibitem{PDG}D.E.Groom {\it et al.} Particle Data Group, Eur. Phys. J. 
{\bf C15}, 1 (2000).



\bibitem{see-saw}
M.~Gell-Mann, P.~Ramond and R.~Slansky,
in \textit{Supergravity}, p.~315, edited by F. van Nieuwenhuizen and D.
  Freedman, North Holland, Amsterdam, 1979\,;
T.~Yanagida,
Proc. of the \textit{Workshop on Unified Theory and the Baryon Number of the
  Universe}, KEK, Japan, 1979\,;
R.N. Mohapatra and G.~Senjanovi{\'c},
Phys. Rev. Lett. \textbf{44}, 912 (1980).

\bibitem{BPet} S.M. Bilenky  and S.T. Petcov,
Rev. Mod. Phys.\textbf{59} (1987) 671.

\bibitem{BGG} S.M.\, Bilenky, C.\, Giunti and
W.\,Grimus. Prog. Part. Nucl. Phys. {\bf 43}, 1 (1999);\,~ hep-ph/9812360.

\bibitem{Kam} Kamiokande collaboration, Y.Fukuda {\it et al.}
Phys. Lett.{\bf 43} (1994) 237.




\bibitem{IMB} IMB collaboration, H. Clark {\it et al.}
Phys. Rev.Lett. {\bf 79} (1997) 345.








\bibitem{Buch}

W. Buchmuller, hep-ph/0204288.

\bibitem{BPin} J. N. Bahcall, M. H. Pinsonneault and S. Basu,
Astrophys. J. {\bf 555}, 990 (2001).

\bibitem{Ortiz}
C.~E.~Ortiz, A. Garcia, R. A. Waltz, M. Bhattacharya and A. K. Komives,
Phys.\ Rev.\ Lett.\  {\bf 85}, 2909 (2000),
nucl-ex/0003006.

\bibitem{BGP}
J. N. Bahcall, M. C. Gonzalez-Carcia, and C.
Pena-Garay, JHEP 0204 (2002) 007; hep-ph/0111150.

\bibitem{Lisi} E. Lisi,
Proceedings of the 20th International Conference on Neutrino
Physics and Astrophysics, {\em Neutrino~2002\/}\,(Munich, Germany,
May 25-30, 2002).

\bibitem{KamL}  KamLAND Collaboration, J. Shirai,
Proceedings of the
20th International Conference on Neutrino
Physics and Astrophysics,
{\em Neutrino~2002\/}\,~(Munich, Germany,
May 25-30, 2002).

\bibitem{CHOOZ} CHOOZ Collaboration,  M.\, Apollonio \textit{et al.},
                 Phys. Lett. B {\bf 466} 415 (1999).
                

\bibitem{PaloV} F. Boehm, J. Busenitz et al.,
               Phys.\ Rev.\ Lett.\  {\bf 84}, 3764 (2000) and
               Phys. Rev. D {\bf 62} (2000) 072002.
 

\bibitem{LSND} LSND Collaboration,
               G. Mills, Proceedings of the
19th International Conference on Neutrino
                Physics and Astrophysics,
{\em Neutrino~2000\/}\,(Sudbury, Canada, June 16-21, 2000).

\bibitem{MiniB} MiniBooNE Collaboration, 
R. Tayloe,
Proceedings of the 20th International Conference on Neutrino
Physics and Astrophysics, {\em Neutrino~2002\/}\,(Munich, Germany
May 25-30, 2002).


\bibitem{BPon}
B. Pontecorvo,
J. Exptl. Theoret. Phys. \textbf{34}, 247 (1958)
[Sov. Phys. JETP \textbf{7}, 172 (1958)]; B. Pontecorvo,
Zh. Eksp. Teor. Fiz. \textbf{53}, 1717 (1967)
[Sov. Phys. JETP \textbf{26}, 984 (1968)].

\bibitem{MNS}
Z. Maki, M. Nakagawa, and S. Sakata,
Prog. Theor. Phys. \textbf{28}, 870 (1962).

\bibitem{Nakaya} T.\,Nakaya, 
Proceedings of the 20th International Conference 
on Neutrino Physics and Astrophysics, {\em Neutrino~2002\/}\,(Munich, Germany
May 25-30, 2002).

\bibitem{Lindner} M.\,Lindner, 
Proceedings of the 20th International Conference 
on Neutrino Physics and Astrophysics, {\em Neutrino~2002\/}\,(Munich, Germany
May 25-30, 2002); hep-ph/0209083.
 
\bibitem{Dydak} F. Dydak, 
Proceedings of the 20th International Conference 
on Neutrino Physics and Astrophysics, {\em Neutrino~2002\/}\,(Munich, Germany
May 25-30, 2002).

\bibitem{Schramm} S.T. Petcov, Phys. Lett. {\bf B 214} 259 (1988);
 X.~Shi and D.N.~ Schramm,
Phys. Lett.  \textbf{ B 283}, 305 (1992).


\bibitem{BG}S.M. Bilenky and C. Giunti, 
Phys.Lett.\textbf{B444}, 379 (1998). 

\bibitem{BOREXINO} BOREXINO Collaboration,
 G. Bellini, Proceedings of the
20th International Conference on Neutrino
                Physics and Astrophysics,
{\em Neutrino~2002\/}\,(Munich, Germany , May 25-30, 2002).

\bibitem{H-M} HEIDELBERG-MOSCOW collaboration, H. V. Klapdor-Kleingrothaus
{\it et al.}, Eur. Phys. J. A \textbf{12}, 147 (2001).

\bibitem{IGEX} IGEX Collaboration,  Aalseth et al., hep-ex/0202026. 

\bibitem{Cremonesi} O. Cremonesi, Proceedings of the
20th International Conference on Neutrino
                Physics and Astrophysics,
{\em Neutrino~2002\/}\,(Munich, Germany , May 25-30, 2002).


\bibitem{BPP} S.M. Bilenky, S.Pascoli and S.T. Petcov,
Phys. Rev.\textbf{D 64} (2001) 053010.



\bibitem{MINOS} MINOS Collaboration,
V. Paolone 
Nucl.Phys.Proc.Suppl.\textbf{100},197 (2001). 


\bibitem{ICARUS} ICARUS Collaboration,
       Proceedings of the "NO-VE International Workshop" Venice 
(Italy, July 2001), (2001) 91  Ed. M. Baldo-Ceolin. 

\bibitem{Itow} Y. Itow \textit{et al.}, hep-ex/0106019.


\end{thebibliography}
\end{document}